\documentclass{article}
\usepackage[utf8]{inputenc}

\usepackage{xcolor}
\usepackage{graphicx}
\usepackage{csquotes}
\usepackage{comment}
\usepackage{booktabs}
\usepackage[]{geometry}
\usepackage{listings}
\usepackage{caption}
\usepackage{subcaption}
\usepackage{titlesec}
\usepackage[titles]{tocloft}
\usepackage{fancyhdr}
\usepackage{url}
\usepackage{lipsum}
\usepackage{amsmath,amssymb,amsfonts}
\usepackage{algorithmic}
\usepackage[]{algorithm2e}
\usepackage{paralist}
\usepackage{datetime2}
\usepackage{soul} 

\newcommand{\jf}[1]{#1} 

\title{The Keyword Explorer Suite: A Toolkit for Understanding Online Populations}
\author{Philip Feldman, Shimei Pan, James R. Foulds}
\date{\today}

\begin{document}
\maketitle

\begin{abstract}
We have developed a set of Python applications that use large language models to identify and analyze data from social media platforms relevant to a population of interest. Our pipeline begins with using OpenAI's GPT-3 to generate potential keywords for identifying relevant text content from the target population. The keywords are then validated, and the content downloaded and analyzed using GPT-3 embedding and manifold reduction. Corpora are then created to fine-tune GPT-2 models to explore latent information \jf{via prompt-based queries}. These tools allow researchers and practitioners to gain valuable insights into population subgroups online.
\end{abstract}

\begin{figure}[h!]
	\hspace*{-2em}  
	\centering
	\begin{subfigure}{.32\textwidth}
		\centering
		\fbox{\includegraphics[height = 12em]{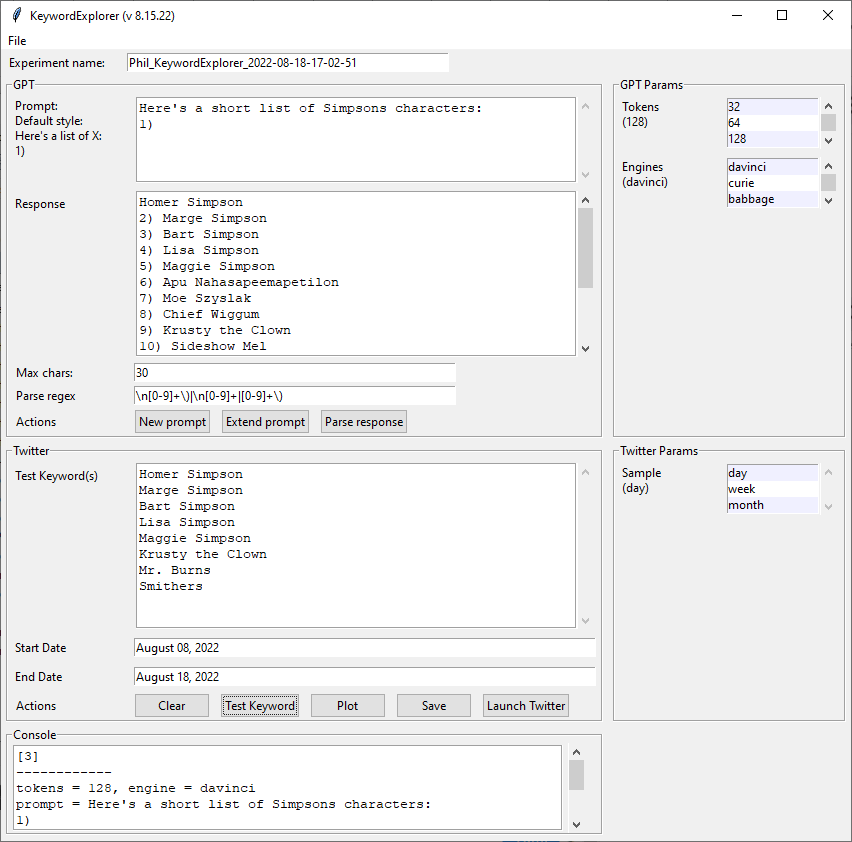}}
		\caption{KeywordExplorer}
		\label{fig:keyword_explorer}
	\end{subfigure}%
	\begin{subfigure}{.47\textwidth}
		\centering
		\fbox{\includegraphics[height = 12em]{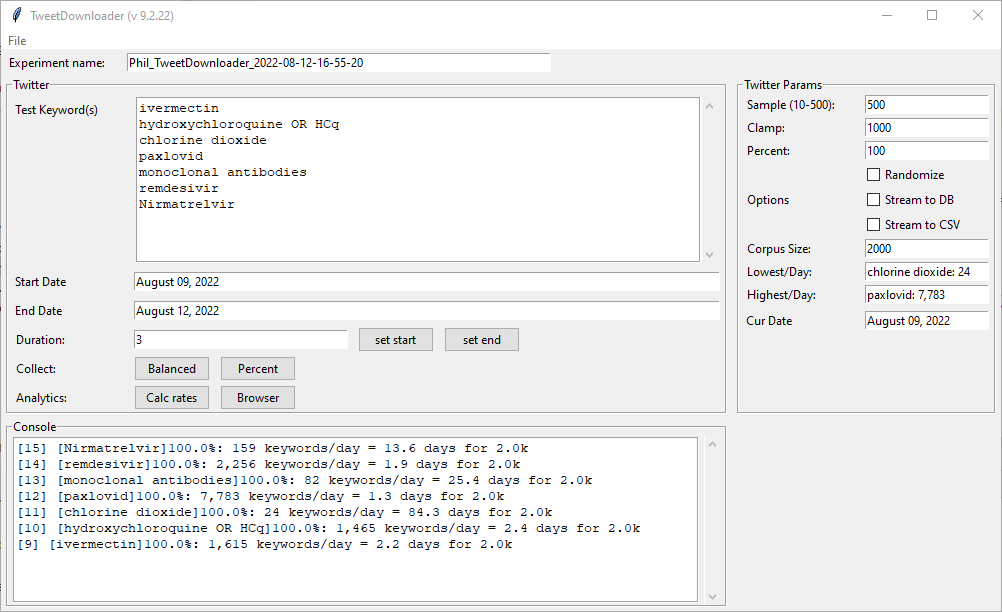}}
		\caption{TweetDownloader}
		\label{fig:tweet_downloader}
	\end{subfigure}%
	\begin{subfigure}{.3\textwidth}
		\centering
		\fbox{\includegraphics[height = 12em]{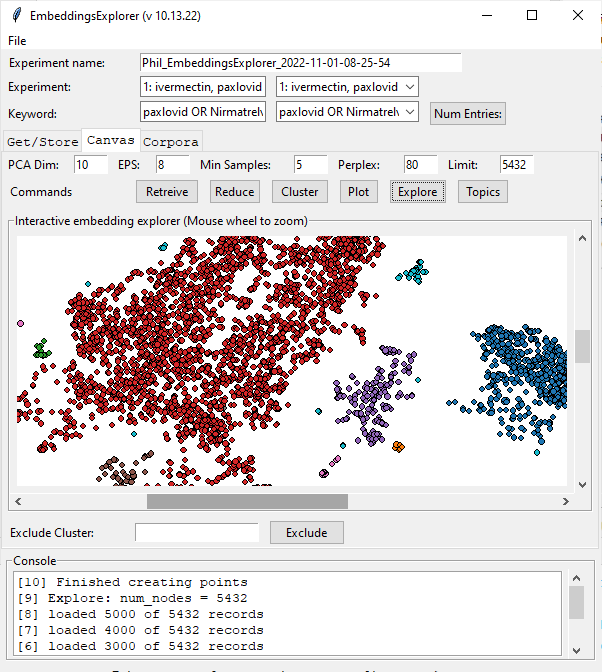}}
		\caption{TweetEmbedExplorer}
		\label{fig:embed_explorer}
	\end{subfigure}%
	\caption{Three Applications from the \jf{Keyword Explorer} Suite.}
	\label{fig:main_apps}
\end{figure}

\section{Introduction}

Imagine a researcher who is trying to understand how a particular population subgroup on social media might react to an event that hasn't happened yet and that they are not currently discussing. This 
\jf{task has a variety of applications} 
such as 
\jf{estimating} 
public opinion, planning marketing campaigns, 
\jf{and} 
identifying potential risks and opportunities. 
\jf{This is not a straightforward task.} 
The researcher must first find the group in question, and they typically do not know the terms that the group uses to describe themselves or their interests. However, large language models like GPT-3 can generate potential \jf{keywords} 
for \jf{identifying content created by} the group  in question since these models are trained on vast amounts of text data and are likely to have ingested interactions from the group.

Once the researcher has identified potential terms, they must be able to verify them against actual social media conversations to see if they are being used by the group or \enquote{hallucinated} by the model. If the terms are relevant, the researcher needs to download, store and clean the appropriate posts, removing any irrelevant or redundant information.

Finally, small language models like GPT-2 can be 
\jf{finetuned} 
 on these posts to create a model \jf{tailored to the group's language and context.}
Because the desired opinions and beliefs are not explicitly in the text, but are latent in the model, the 
model can be used to 
\jf{generate artificial text simulating}  
how the population might react to the hypothetical event.

This scenario illustrates how language models can 
identify and understand population subgroups on social media, even when the desired information is not explicitly present in the text. Such models can provide valuable insights into the thoughts and behaviors of these groups, allowing researchers to make more informed decisions and predictions.


To address this 
\jf{scenario}, we have developed a 
\jf{Python toolkit for using} 
large language models to identify and analyze relevant data from social media platforms. 
\jf{Our} applications 
\jf{facilitate}  
understanding and analyzing population subgroups online, enabling researchers and practitioners to gain insights that would not be possible through traditional methods.



\section{Background}

\emph{Query term identification} is 
\jf{the process of identifying relevant terms and phrases for describing} 
a particular topic or concept. 
However, this process is often ad-hoc and deeply reflects the researcher's awareness and bias~\cite{king2017computer, buntain2018sampling}. Other approaches rely on query logs  and cannot be used for recommending important words based on a domain where the user has little prior knowledge or experience~\cite{felser2022recommendation}. This can lead to a lack of consistency and reproducibility. 

%


In recent years, there have been 
attempts to create tools 
\jf{to} 
help researchers determine the optimal keywords for search in social media in 
disciplines 
such as information retrieval. These tools often use various techniques, such as natural language processing~\cite{eads2021separating} and machine learning~\cite{mauriello2018smidgen}, to analyze large datasets and identify relevant keywords.

For example, some researchers have used topic modeling algorithms, such as Latent Dirichlet Allocation (LDA), to identify the most common themes and topics in a dataset~\cite{feizollah2019halal}. These themes can then be used as keywords to search for relevant content. Other researchers have used sentiment analysis to identify the sentiment associated with specific keywords, which can be useful for understanding how different groups of people feel about a particular topic~\cite{mauriello2018smidgen}.

There have also been efforts to develop keyword generation tools specifically for social media platforms, such as Twitter~\cite{biswas2018graph}. These tools often rely on machine learning algorithms to identify patterns in the language and behavior of users, and use this information to suggest relevant keywords for search~\cite{kwok2021tweet}.

\jf{While} these tools have had some success in helping researchers identify relevant keywords, 
many of these tools rely on supervised learning techniques, which require large amounts of labeled data for training~\cite{kadhim2019survey}. This can be difficult to obtain, particularly in domains where there is limited available data or where the language used is highly specialized. Additionally, these tools \jf{(e.g. LDA)} may be limited in their ability to capture the complexity and nuance of human language and behavior, which can be important for understanding social media conversations and other phenomena.

\jf{Alternatively, transformer} 
language models such as BERT and GPT are trained on vast amounts of data from a wide range of sources, including books, articles, and social media posts, and as a result, they have a broad understanding of language and context. This can make them useful for sociology research, such as \jf{addressing query term identification by} generating lists of slang terms or other specialized vocabulary that may be difficult to find using other means~\cite{feldman2021analyzing}.

These large language models could potentially further be used to analyze the retrieved data, but they may not be tailored to the specific needs and concerns of examining a particular subgroup. To address this, smaller language models such as GPT-2 can be quickly fine-tuned on corpora that are specific to a 
\jf{target} 
domain~\cite{feldman2020navigating}. This allows the models to capture the language and context of particular social subgroups, enabling a new form of computational sociology. By repeatedly prompting these finetuned models to produce a large volume of responses, researchers can gain insight into the language and behaviors of these groups in a more nuanced and detailed manner than \jf{via} traditional means~\cite{dant2022polling}.

\section{Application Pipeline Overview}


The 
\jf{Keyword Explorer Suite} 
\jf{is a toolkit for understanding online populations, consisting of} 
a set of Python applications that work together in a pipeline, where each app produces outputs that are used in subsequent applications. The suite includes a graphical user interface (GUI) that allows the user to explore the data in an interactive environment. 

The pipeline consists of:  \textit{keyword exploration and validation}, continues with \textit{data collection}, then \textit{data cleaning and refinement}, and concludes with \textit{model training and exploration}. These parts of the pipeline are discussed in detail below: 

\subsection{Keyword Exploration and Validation}
Keyword exploration uses the \textit{KeywordExplorer} app (Figure~\ref{fig:keyword_explorer}), which lets the user prompt the GPT-3 using the OpenAI API to produce lists of keywords. Prompts are generally of the form: 
\jf{\enquote{\textit{xxxx: <newline> 1)}}. 
Here, \textit{<newline>} is a line break and}   \textit{xxxx} is a string such as \textit{List slang terms that have been used to refer to COVID-19:}, or \textit{Create a list of hashtags that are important on Black Twitter}. As an illustrative example we will be using \textit{Here's a short list of exotic pets}. Each time the model is prompted, slightly different responses will be generated, and can be evaluated. For this \jf{query,} 
the responses started with \enquote{1) Bats, 2) Monkeys, 3) Snakes,} and went to \enquote{9) Tarantulas, 10) Scorpions, and 11) Sugar Gliders.}

When prompted in this form, such responses from the GPT-3 are easily parsed using a regex. Once applied, an unadorned list is produced that can be passed to Twitter or Wikipedia to evaluate the \jf{keywords based on the} usage of each keyword over time. 
\jf{With the default parameters we retrieve count data for} 
10 days in the past to the current date, though these are easily edited. Submitting these words to Twitter returns the total usage over that period (Dec. 17-27, 2022) to be Monkeys (36,772), Snakes (29,830), Bats (21,156), Alligators (3,258), Tarantulas (689), and Sugar Gliders (196). To further evaluate the applicability of the keywords, the app can launch a series of tabs in the default browser for each of the keywords across the defined timeframe. This can \jf{help to} find times when the context switches, as in the case for \enquote{Birds} during 2014, when the \textit{Baltimore Orioles} baseball team, referred to as \enquote{The Birds,} had a particularly good season.


\subsection{Data Collection}
Once the keywords are validated, they are passed into the \textit{TweetDownloader} app (Figure~\ref{fig:tweet_downloader}). This app allows the user to submit the keywords individually to the Twitter API and sample the relevant 
\jf{tweets} 
over a defined period. The number of 
\jf{tweets} 
per keyword per day can be adjusted so that they are the same for each day or proportional with respect to the largest number of tweets for any particular keyword.
The \textit{TweetDownloader} app also allows the user to also filter for language, location, and other criteria. The results are saved to a MariaDB relational database, which allows sophisticated queries and analysis.  The database stores data and time information, which supports downstream chronological sampling. Users can apply daily and overall limits for each keyword \jf{corpus}. 

\subsection{Data Cleaning and Refinement}
\textit{TweetEmbedExplorer} (Figure~\ref{fig:embed_explorer}) filters, analyzes, and augments tweet information. Augmented information can then be used to create a train/test corpus for finetuning language models such as the GPT-2. For any keyword, user information associated with the ID of each post can be collected and placed in its own table in the database, allowing more sophisticated queries. Using the OpenAI \textit{text embeddings} API, the embeddings for each tweet can be stored in the database. These can be projected to 2D using a combination of PCA and T-SNE. Once dimension reduction has been performed, clustering can be performed interactively using DBSCAN~\cite{delise2021data}, 
%
\jf{and outlier clusters can be marked \enquote{excluded.}} 

\jf{The next step} 
is to produce a 
\jf{corpus} 
for finetuning a GPT-2 model. 
\jf{Finetuned} models have been shown to 
accurately predict, for example, the vegetarian preferences of Yelp reviewers when all vegetarian data has been excluded from the test/train data~\cite{dant2022polling}. 
\jf{This} means that a model \jf{finetuned} on a set of tweets that may not contain the explicit information about a certain subject 
\jf{may} 
still be able to generate tweets that are likely to contain that information \jf{in some cases}.

We use a process we call \enquote{\emph{meta wrapping}} that creates text that has 
\jf{metadata} 
in it in addition to the 
tweet\jf{, e.g.:}

\noindent
{\small 
\texttt{[[text: I know I'm not the prettiest dog but my love for you is unconditional always because I have a beautiful heart and soul  || created: 2022-12-27 07:10:25 || location: USA || probability: twenty]]}}

Each entry in the string is wrapped by \texttt{[[} and \texttt{]]}. All elements within the meta wrapped string are separated by \texttt{||}. The \enquote{text:} element precedes the posting, the \enquote{created:} tag precedes the data the post was created, and the \enquote{location:} tag contains the coordinates of the tweet or the poster if available. For model \jf{convergence} validation, there is a substring of the form \enquote{probability: xxxx}, where xxxx can be \enquote{ten}, \enquote{twenty}, \enquote{thirty}, or \enquote{forty.} The likelihood that a line will have the appropriate string reflects the probability of the random number generator hitting that value. 
\jf{As a sanity check, if the generated data does not match these percentages, insufficient finetuning can clearly be diagnosed.} 

\subsection{Model Training and Exploration}
\begin{figure}[t]
    \centering
    \fbox{\includegraphics[width = \textwidth]{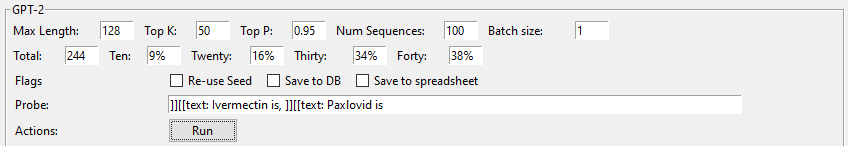}}
    \caption{\jf{ModelExplorer:} GPT Parameters 
     \jf{and the Results of the Meta Wrapping Probability Sanity Check.}
    }
    \label{fig:model_explorer_params}
\end{figure}

\textit{ModelExplorer}  is a tool that allows users to \jf{interactively} explore finetuned GPT-2 models \jf{(Figure~\ref{fig:model_explorer_params})}. These models are finetuned on tweet corpora generated from TweetEmbedExplorer. 
It 
\jf{allows} 
the user to provide a set of comma-separated probes along with model hyperparameters and then generate a series of predictions from the model. 
Text that is output from the model is parsed, displayed in the text output area of the tool, and if desired, stored in the database for further analysis. 
%
\jf{In the example in Figure~\ref{fig:model_explorer_params}, probes regarding Ivermectin and Paxlovid were used. The results show that there was a maximum deviation of 4\% with respect to the target percentages, so the finetuning sanity check was passed.}

\section{Conclusions}
\jf{We} presented 
\jf{a Python toolkit}  
\jf{for gaining insight} 
into population subgroups.\footnote{Full code available at \url{https://github.com/pgfeldman/KeywordExplorer}.} 
Our pipeline begins with 
\jf{using} 
large language models to generate potential keywords for exploring population subgroups, which are then validated using historical usage trends on social media platforms. The resulting data is then analyzed and clustered using a combination of GPT-3 embeddings and manifold reduction. Finally, the posts are used to generate corpora that can be used to fine-tune language models and identify latent information. 
\jf{We plan to use our toolkit to study COVID-19 racism.} 


\section*{Acknowledgements}
We would like to thank OpenAI for giving us permission to use the GPT-3 in creepy in ways that are not really per their policy, HuggingFace for their wonderful ecosystem and API, and Twitter for (as of this writing) not falling apart.


\begin{thebibliography}{10}

\bibitem{biswas2018graph}
Saroj~Kr Biswas, Monali Bordoloi, and Jacob Shreya.
\newblock A graph based keyword extraction model using collective node weight.
\newblock {\em Expert Systems with Applications}, 97:51--59, 2018.

\bibitem{buntain2018sampling}
Cody Buntain, Erin McGrath, and Brandon Behlendorf.
\newblock Sampling social media: Supporting information retrieval from
  microblog data resellers with text, network, and spatial analysis.
\newblock In {\em Proceedings of the 51st Hawaii International Conference on
  System Sciences}, 2018.

\bibitem{delise2021data}
Timothy DeLise.
\newblock Data segmentation via t-sne, dbscan, and random forest.
\newblock In {\em Intelligent Computing}, pages 139--151. Springer, 2021.

\bibitem{eads2021separating}
Alicia Eads, Alexandra Schofield, Fauna Mahootian, David Mimno, and Rens
  Wilderom.
\newblock Separating the wheat from the chaff: A topic and keyword-based
  procedure for identifying research-relevant text.
\newblock {\em Poetics}, 86:101527, 2021.

\bibitem{feizollah2019halal}
Ali Feizollah, Sulaiman Ainin, Nor~Badrul Anuar, Nor Aniza~Binti Abdullah, and
  Mohamad Hazim.
\newblock Halal products on twitter: Data extraction and sentiment analysis
  using stack of deep learning algorithms.
\newblock {\em IEEE Access}, 7:83354--83362, 2019.

\bibitem{feldman2020navigating}
Philip Feldman.
\newblock Navigating language models with synthetic agents.
\newblock {\em arXiv preprint arXiv:2008.04162}, 2020.

\bibitem{dant2022polling}
Philip Feldman, Aaron Dant, James~R Foulds, and Shimei Pan.
\newblock Polling latent opinions: A method for computational sociolinguistics
  using transformer language models.
\newblock {\em arXiv preprint arXiv:2204.07483}, 2022.

\bibitem{feldman2021analyzing}
Philip Feldman, Sim Tiwari, Charissa~SL Cheah, James~R Foulds, and Shimei Pan.
\newblock Analyzing covid-19 tweets with transformer-based language models.
\newblock {\em arXiv preprint arXiv:2104.10259}, 2021.

\bibitem{felser2022recommendation}
Jenny Felser, Jian Xi, Christoph Demus, Dirk Labudde, and Michael Spranger.
\newblock Recommendation of query terms for colloquial texts in forensic text
  analysis.
\newblock {\em INFORMATIK 2022}, 2022.

\bibitem{kadhim2019survey}
Ammar~Ismael Kadhim.
\newblock Survey on supervised machine learning techniques for automatic text
  classification.
\newblock {\em Artificial Intelligence Review}, 52(1):273--292, 2019.

\bibitem{king2017computer}
Gary King, Patrick Lam, and Margaret~E Roberts.
\newblock Computer-assisted keyword and document set discovery from
  unstructured text.
\newblock {\em American Journal of Political Science}, 61(4):971--988, 2017.

\bibitem{kwok2021tweet}
Stephen Wai~Hang Kwok, Sai~Kumar Vadde, and Guanjin Wang.
\newblock Tweet topics and sentiments relating to covid-19 vaccination among
  australian twitter users: machine learning analysis.
\newblock {\em Journal of Medical Internet Research}, 23(5):e26953, 2021.

\bibitem{mauriello2018smidgen}
Matthew~Louis Mauriello, Cody Buntain, Brenna McNally, Sapna Bagalkotkar,
  Samuel Kushnir, and Jon~E Froehlich.
\newblock {SMIDGen}: An approach for scalable, mixed-initiative dataset
  generation from online social networks.
\newblock {\em HCIL Tech Reports}, pages 2018--01, 2018.

\end{thebibliography}

\end{document}